\documentclass[conference,romanappendices,10pt]{IEEEtran}	

\IEEEoverridecommandlockouts

\usepackage{amsmath,amssymb,amsthm,cite,color,graphicx,microtype,url}
\usepackage[USenglish]{babel}
\usepackage[utf8]{inputenc} 
\usepackage[T1]{fontenc}

{}
{}
{}
{\newtheorem{proposition}{Proposition}}
{}
{}
{}

\renewcommand{\a}{\mathbf{a}}

\newcommand{\e}{\mathbf{e}}

\newcommand{\h}{\mathbf{h}}

\newcommand{\s}{\mathbf{s}}

\newcommand{\w}{\mathbf{w}}
\newcommand{\x}{\mathbf{x}}

\newcommand{\0}{\mathbf{0}}

\newcommand{\E}{\mathbf{E}}

\renewcommand{\H}{\mathbf{H}}
\newcommand{\I}{\mathbf{I}}

\newcommand{\W}{\mathbf{W}}

\newcommand{\mub}{\boldsymbol{\mu}}

\newcommand{\Deltab}{\mathbf{\Delta}}

\newcommand{\Sigmab}{\mathbf{\Sigma}}
\newcommand{\Upsilonb}{\boldsymbol{\Upsilon}}

\newcommand{\setC}{\mathcal{C}}

\newcommand{\setN}{\mathcal{N}}

\newcommand{\setW}{\mathcal{W}}

\newcommand{\Compl}{\mbox{$\mathbb{C}$}}

\newcommand{\argmin}{\operatornamewithlimits{argmin}}
\newcommand{\argmax}{\operatornamewithlimits{argmax}}
\newcommand{\blkdiag}{\mathrm{blkdiag}}

\newcommand{\Exp}{\mathbb{E}}
\newcommand{\herm}{\mathrm{H}}

\newcommand{\tran}{\mathrm{T}}

\newcommand{\rmF}{\mathrm{F}}
\newcommand{\rmh}{\mathrm{h}}
\newcommand{\fb}{\mathrm{fb}}
\newcommand{\gr}{\textnormal{\tiny{GR}}}
\newcommand{\lr}{\textnormal{\tiny{LR}}}
\newcommand{\na}{\textnormal{\tiny{NA}}}
\newcommand{\tx}{\mathrm{tx}}
\newcommand{\rzf}{\mathrm{rzf}}

\newcommand\narrowstyleb{\SetTracking{encoding=*}{-11}\lsstyle}

\title{Cooperative MIMO Precoding with Distributed CSI: A Hierarchical Approach}
\author{
\IEEEauthorblockN{Italo Atzeni and David Gesbert}
\IEEEauthorblockA{EURECOM, Communication Systems Department, Sophia Antipolis, France}
Emails: \{italo.atzeni, david.gesbert\}@eurecom.fr
\thanks{This work was supported by the European Research Council under the Horizon 2020 Programme (ERC 670896 PERFUME).}}

\begin{document}

\maketitle

\begin{abstract}
The problem of network multiple-input multiple-output precoding under distributed channel state information is~a notoriously challenging question, for which optimal solutions with reasonable complexity remain elusive. In this context, we~assess the value of hierarchical information exchange, whereby an~order is established among the transmitters (TXs) in such a way that~a given TX has access not only to its local channel estimate~but~also to the estimates available at the less informed TXs. Assuming regularized zero forcing (RZF) precoding at the TXs, we propose naive, locally robust, and globally robust suboptimal strategies~for the joint precoding design. Numerical results show that hierarchical information exchange brings significant performance gains, with the locally and globally robust algorithms performing remarkably close to the optimal~RZF~strategy.~Lastly,~the cost~of hierarchical information exchange relative to the cooperation {\narrowstyleb gain is examined and the optimal tradeoff is numerically evaluated.}
\end{abstract}



\section{Introduction} \label{sec:INTRO}

Network multiple-input multiple-output (MIMO) systems, whereby distributed transmitters (TXs) sharing user data symbols and channel state information (CSI) cooperatively serve several receivers (RXs) by cooperatively designing their downlink precoding strategy, are regarded as a promising solution to enhance data rates and to meet the quality-of-service requirements of future cellular networks \cite{Ges10}. A practical limitation of decentralized network MIMO systems, which makes the joint precoding optimization challenging, is that the CSI is actually known imperfectly and differently across the TXs due to limited and uneven feedback \cite{deK12}: this occurs, for instance, when the TXs are not connected to a perfect backhaul or are mounted on mobile devices such as vehicles or drones \cite{Moz15}. Under such a \textit{distributed CSI} (D-CSI) setting, each TX needs to design its precoding strategy solely on the basis of its local CSI without any further information exchange with the other TXs. This problem falls into the category of so-called \textit{team decision problems} \cite{Rad62}, where multiple decentralized decision makers aim at coordinating their strategies to maximize the system-level performance while not being able to accurately predict the actions taken by the other decision makers.

Under centralized CSI, the network MIMO TXs can be virtually combined into a unique antenna array serving the RXs in a multi-antenna broadcast channel fashion, for which there is a large body of literature dealing with robust precoding in presence of CSI imperfections (see, e.g., \cite{She08,Vuc09}). On the other hand, fewer results are available for the D-CSI case. Among these, the work in \cite{deK16} proposes a robust distributed precoding method that relies on the quantization of the CSI space to enforce coordination between the TXs. In \cite{Lak16}, D-CSI arises from combining periodical feedback via backhaul links, equal for all TXs, with local CSI exchanges from neighboring RXs, generating partially new local CSI at a given TX between backhaul updates. Furthermore, \cite{deK14} proposes a D-CSI structure where the TXs are ordered by increasing level of CSI, i.e., in such a way that a given TX has access not only to its local CSI but also to the CSI available at the less informed TXs.

In this paper, we formulate the general joint precoding optimization problem under D-CSI as a team decision problem and particularize it to a \textit{deterministically hierarchical D-CSI} scenario \cite{deK14}. Hierarchical D-CSI can be enforced by a suitable information exchange mechanism between the TXs at a certain signaling/power cost: here, we show how such a hierarchical information exchange can be leveraged to yield implementable and efficient distributed precoding solutions. In particular, restricting the structure of the precoding strategies to \textit{regularized zero forcing} (RZF) precoding \cite{Pee05} and considering the ergodic sum rate as performance metric, we propose~\textit{naive}, \textit{locally robust}, and \textit{globally robust} suboptimal strategies (in increasing order of both performance and computational complexity) for the joint precoding design. {\narrowstyleb Numerical results show that the deterministically hierarchical D-CSI configuration yields significant gains over the classical non-hierarchical D-CSI counterpart, with the locally and globally robust algorithms performing remarkably close to the optimal RZF strategy. Lastly, the cost of hierarchical information exchange is examined and is shown to be outweighed by the resulting cooperation gain.}

\begin{figure*}[!]
\addtocounter{equation}{+5}
\begin{align} \label{eq:W_star_gen}
\begin{array}{ccl}
\vspace{1mm}
\displaystyle \W_{\star}^{(n)} \triangleq & \displaystyle \hspace{-2mm} \argmax_{\W^{(n)}} & \displaystyle \hspace{-2mm} \Exp_{\{ \hat{\H}^{(\ell)} \}_{\ell \neq n} | \hat{\H}^{(n)}} \bigg[ \max_{\{ \W^{(\ell)} \}_{\ell \neq n}} \Exp_{\H | \hat{\H}^{(n)}} \bigg[ R \Big( \H, \W^{(n)}(\hat{\H}^{(n)}), \big\{ \W^{(\ell)}(\hat{\H}^{(\ell)}) \big\}_{\ell \neq n} \Big) \bigg] \bigg] \\
& \hspace{-2mm} \mathrm{s.t.} & \hspace{-2mm} \big\| \W^{(\ell)} (\hat{\H}^{(\ell)}) \big\|_{\rmF}^{2} \leq P_{\ell}, \qquad \ell = 1, \ldots, N
\end{array}
\end{align}
\hrulefill
\addtocounter{equation}{+4}
\begin{align} \label{eq:W_star}
\hspace{-2mm} \begin{array}{ccl}
\vspace{1mm}
\displaystyle \W_{\star,\rmh}^{(n)} \triangleq & \displaystyle \hspace{-2mm} \argmax_{\W^{(n)}} & \displaystyle \hspace{-2mm} \Exp_{\{ \hat{\H}^{(\ell)} \}_{\ell=n+1}^{N} | \hat{\H}^{(n)}} \bigg[ \max_{\{ \W^{(\ell)} \}_{\ell=n+1}^{N}} \Exp_{\H | \hat{\H}^{(n)}} \bigg[ R \Big( \H, \big\{ \W_{\star,\rmh}^{(\ell)} \big\}_{\ell=1}^{n-1}, \W^{(n)}(\hat{\H}^{(n)}), \big\{ \W^{(\ell)}(\hat{\H}^{(\ell)}) \big\}_{\ell=n+1}^{N} \Big) \bigg] \bigg] \\
& \hspace{-2mm} \mathrm{s.t.} & \hspace{-2mm} \big\| \W^{(\ell)} (\hat{\H}^{(\ell)}) \big\|_{\rmF}^{2} \leq P_{\ell}, \qquad \ell = n, \ldots, N
\end{array}
\end{align}
\hrulefill
\vspace{-1.4mm}
\end{figure*}

\section{System Model} \label{sec:SM}

Let us consider a network MIMO system where $N$ distributed multi-antenna TXs cooperatively serve $K$ single-antenna RXs in the downlink. Each TX~$n$ is equipped with $M_{n}$ antennas and the total number of transmit antennas among the $N$ TXs is $M \triangleq \sum_{n=1}^{N} M_{n}$. We use $\h_{k n} \in \Compl^{M_{n} \times 1}$, $\h_{k} \triangleq [\h_{k 1}^{\tran} \ldots \h_{k N}^{\tran}]^{\tran} \in \Compl^{M \times 1}$, and $\H \triangleq [\h_{1} \ldots \h_{K}] \in \Compl^{M \times K}$ to denote the channel vector between TX~$n$ and RX~$k$, the channel vector between the $N$ TXs and RX~$k$, and the channel matrix between the $N$ TXs and the $K$ RXs, respectively. Furthermore, we assume that $\h_{kn} \sim \setC \setN (\0, \Sigmab_{kn})$, with $\Sigmab_{kn} \in \Compl^{M_{n} \times M_{n}}$ being the covariance matrix of $\h_{kn}$: hence, it follows that $\h_{k} \sim \setC \setN (\0, \Sigmab_{k})$, with $\Sigmab_{k} \triangleq \blkdiag (\Sigmab_{k1}, \ldots, \Sigmab_{kN}) \in \Compl^{M \times M}$ being the covariance matrix of $\h_{k}$.

Let $\W \! \in \! \Compl^{M \times K} \!$ denote the multiuser precoding matrix given by \addtocounter{equation}{-11}
\begin{align}
\W \triangleq [\w_{1} \ldots \w_{K}] = \begin{bmatrix}
\W^{(1)} \vspace{-2mm} \\
\vdots \\
\W^{(N)}
\end{bmatrix}
\end{align}
where $\w_{k} \in \Compl^{M \times 1}$ is the beamforming vector used by the $N$ TXs to serve RX~$k$ and $\W^{(n)} \in \Compl^{M_{n} \times K}$ is the precoding submatrix used by TX~$n$; a per-TX power constraint is assumed such that $\|\W^{(n)} \|_{\rmF}^{2} \leq P_{n}$. The receive signal at RX~$k$ is then expressed as
\begin{align} \label{eq:y_k}
y_{k} \triangleq \h_{k}^{\herm} \x + z_{k}
\end{align}
where $\x \in \Compl^{M \times 1}$ is the transmit signal obtained from the user data symbol vector $\s \triangleq [s_{1} \ldots s_{K}]^{\tran} \in \Compl^{K \times 1}$ as
\begin{align}
\x \triangleq \W \s = \sum_{k=1}^{K} \w_{k} s_{k}
\end{align}
and $z_{k} \sim \setC \setN (0,\sigma^{2})$ is the noise at RX~$k$.
Finally, the sum rate of the network MIMO system is given by
\begin{align} \label{eq:R}
R(\H, \W) \triangleq \sum_{k=1}^{K} \log_{2} \bigg( 1 + \frac{|\h_{k}^{\herm} \w_{k}|^{2}}{\sum_{j \neq k} |\h_{k}^{\herm} \w_{j}|^{2} + \sigma^{2}} \bigg).
\end{align}

\vspace{1mm}

\section{Distributed CSI Model} \label{sec:DCSI}

In practice, not only is the channel matrix known imperfectly but also differently across the network nodes due to limited and uneven feedback. In this paper, we thus consider a D-CSI scenario \cite{deK12}, where each TX $n$ has a different estimate of the channel matrix $\H$,\footnote{Even though the CSI is distributed, it is still reasonable to assume that the user data symbol vector $\s$ is perfectly known at all TXs.} denoted by $\hat{\H}^{(n)} = [\hat{\h}_{1}^{(n)} \ldots \hat{\h}_{K}^{(n)}] \in \Compl^{M \times K}$. The imperfect CSI at TX $n$ is modeled as
\begin{align} \label{eq:csi}
\hat{\H}^{(n)} \triangleq \sqrt{1 - \epsilon_{n}^{2}} \H + \epsilon_{n} \E^{(n)}
\end{align}
where $\epsilon_{n} \in [0, 1]$ describes the quality of the channel estimation and $\E^{(n)} \triangleq [\e_{1}^{(n)} \ldots \e_{K}^{(n)}] \in \Compl^{M \times K}$ is the error matrix, where $\e_{k}^{(n)} \sim \setC \setN (0, \Upsilonb^{(n)})$, $\forall k = 1, \ldots, K$, with $\Upsilonb^{(n)} \in \Compl^{M \times M}$ being the error covariance matrix of TX~$n$. 

Hence, in a D-CSI scenario, it is meaningful to formulate a team decision problem (see \cite{Rad62}) where each TX~$n$ computes its precoding submatrix with the objective of maximizing the ergodic sum rate given the local channel estimate $\hat{\H}^{(n)}$, as shown in \eqref{eq:W_star_gen} at the top of the page, where we have expressed the precoding submatrix computed by each TX as a function of its local channel estimate. The conditional distributions of $\H | \hat{\H}^{(n)}$ and $\{ \hat{\H}^{(\ell)} \}_{\ell \neq n} | \hat{\H}^{(n)}$, by which TX $n$ can make a prediction on the real channel matrix and on the CSI available at the other TXs, respectively, are derived in the following~proposition.

\begin{proposition} \label{prop:cond}
Given the unconditional channel $\h_{k} \sim \setC \setN (\0, \Sigmab_{k})$ and the channel estimation model in \eqref{eq:csi}, the following hold:
\begin{itemize}
\item[\textrm{i)}] The channel $\h_{k}$ conditioned on the channel estimate $\hat{\h}_{k}^{(n)}$ is distributed as $\h_{k} | \hat{\h}_{k}^{(n)} \sim \setC \setN (\mub_{k}^{(n)}, \Sigmab_{k}^{(n)})$, with \addtocounter{equation}{+1}
\begin{align}
\label{eq:mu_k_n} \hspace{-7mm} \mub_{k}^{(n)} & \! \triangleq \sqrt{1 - \epsilon_{n}^{2}} \Sigmab_{k} \big( (1 - \epsilon_{n}^{2}) \Sigmab_{k} + \epsilon_{n}^{2} \Upsilonb^{(n)} \big)^{-1} \hat{\h}_{k}^{(n)}, \\
\label{eq:Sigma_k_n} \hspace{-7mm} \Sigmab_{k}^{(n)} & \! \triangleq \Sigmab_{k} - \! (1 - \epsilon_{n}^{2}) \Sigmab_{k} \big( (1 - \epsilon_{n}^{2}) \Sigmab_{k} + \epsilon_{n}^{2} \Upsilonb^{(n)} \big)^{-1} \Sigmab_{k}.
\end{align}
\item[\textrm{ii)}] The channel estimate $\hat{\h}_{k}^{(\ell)}$ conditioned on the channel estimate $\hat{\h}_{k}^{(n)}$ is distributed as $\hat{\h}_{k}^{(\ell)} | \hat{\h}_{k}^{(n)} \sim \setC \setN (\mub_{k}^{(\ell | n)}, \Sigmab_{k}^{(\ell | n)})$, with
\begin{align}
\label{eq:mu_k_ln} \mub_{k}^{(\ell| n)} & \triangleq \sqrt{1 - \epsilon_{\ell}^{2}} \mub_{k}^{(n)}, \\
\label{eq:Sigma_k_ln} \Sigmab_{k}^{(\ell | n)} & \triangleq (1 - \epsilon_{\ell}^{2}) \Sigmab_{k}^{(n)} + \epsilon_{\ell}^{2} \Upsilonb^{(\ell)}
\end{align}
with $\mub_{k}^{(n)}$ and $\Sigmab_{k}^{(n)}$ defined in \eqref{eq:mu_k_n} and \eqref{eq:Sigma_k_n}, respectively.
\end{itemize}
\end{proposition}
\noindent In Proposition~\ref{prop:cond}, $\mub_{k}^{(n)}$ (resp. $\mub_{k}^{(\ell| n)}$) is computed as the minimum mean square error (MMSE) estimate of $\h_{k} | \hat{\h}_{k}^{(n)}$ (resp. $\hat{\h}_{k}^{(\ell)} | \hat{\h}_{k}^{(n)}$), whereas $\Sigmab_{k}^{(n)}$ (resp. $\Sigmab_{k}^{(\ell | n)}$) is the corresponding MMSE covariance matrix. In the rest of the paper, we assume that the channel covariance matrices $\{ \Sigmab_{k} \}_{k = 1}^{K}$, the error covariance matrices $\{ \Upsilonb^{(n)} \}_{n=1}^{N}$, and the coefficients $\{ \epsilon_{n} \}_{n=1}^{N}$ are perfectly known across the network\footnote{Observe that this is a realistic assumption since these parameters depend on long-term statistics.} so that each TX~$n$ can derive the conditional distributions of $\H | \hat{\H}^{(n)}$ and $\{ \H^{(\ell)} \}_{\ell \neq n} | \hat{\H}^{(n)}$ using \eqref{eq:mu_k_n}--\eqref{eq:Sigma_k_n} and \eqref{eq:mu_k_ln}--\eqref{eq:Sigma_k_ln}, respectively.

\begin{figure*}[t!]
\addtocounter{equation}{+2}
\begin{align}
\nonumber \alpha_{\star,\rmh}^{(n)} \triangleq \argmax_{\alpha^{(n)} \in [0,1]} \Exp_{\{ \hat{\H}^{(\ell)} \}_{\ell=n+1}^{N} | \hat{\H}^{(n)}} \bigg[ \max_{\{ \alpha^{(\ell)} \in [0,1] \}_{\ell=n+1}^{N}} \Exp_{\H | \hat{\H}^{(n)}} \bigg[ R \Big( & \H, \big\{ \W_{\rzf}^{(\ell)}(\hat{\H}^{(\ell)}, \alpha_{\star,\rmh}^{(\ell)}) \big\}_{\ell=1}^{n-1}, \\
\label{eq:alpha_star} & \W_{\rzf}^{(n)}(\hat{\H}^{(n)}, \alpha^{(n)}), \big\{ \W_{\rzf}^{(\ell)}(\hat{\H}^{(\ell)}, \alpha^{(\ell)}) \big\}_{\ell=n+1}^{N} \Big) \bigg] \bigg]
\end{align}
\hrulefill
\begin{align} \label{eq:alpha_na}
\alpha_{\na,\rmh}^{(n)} & \triangleq \argmax_{\alpha^{(n)} \in [0,1]} R \Big( \hat{\H}^{(n)}, \big\{ \W_{\rzf}^{(\ell)}(\hat{\H}^{(\ell)}, \alpha_{\na,\rmh}^{(\ell)}) \big\}_{\ell=1}^{n-1}, \W_{\rzf}^{(n)}(\hat{\H}^{(n)}, \alpha^{(n)}), \big\{ \W_{\rzf}^{(\ell)}(\hat{\H}^{(n)}, \alpha^{(n)}) \big\}_{\ell=n+1}^{N} \Big) \\
\label{eq:alpha_lr} \alpha_{\lr,\rmh}^{(n)} & \triangleq \argmax_{\alpha^{(n)} \in [0,1]} \Exp_{\H | \hat{\H}^{(n)}} \Big[ R \Big( \H, \big\{ \W_{\rzf}^{(\ell)}(\hat{\H}^{(\ell)}, \alpha_{\lr,\rmh}^{(\ell)}) \big\}_{\ell=1}^{n-1}, \W_{\rzf}^{(n)}(\hat{\H}^{(n)}, \alpha^{(n)}), \big\{ \W_{\rzf}^{(\ell)}(\hat{\H}^{(n)}, \alpha^{(n)}) \big\}_{\ell=n+1}^{N} \Big) \Big] \\
\label{eq:alpha_gr} \alpha_{\gr,\rmh}^{(n)} & \triangleq \argmax_{\alpha^{(n)} \in [0,1]} \Exp_{\{ \hat{\H}^{(\ell)} \}_{\ell=n+1}^{N} | \hat{\H}^{(n)}} \! \bigg[ \Exp_{\H | \hat{\H}^{(n)}} \! \Big[ R \Big( \H, \! \big\{ \W_{\rzf}^{(\ell)}(\hat{\H}^{(\ell)}, \alpha_{\gr,\rmh}^{(\ell)}) \big\}_{\ell=1}^{n-1}, \! \W_{\rzf}^{(n)}(\hat{\H}^{(n)}, \alpha^{(n)}), \! \big\{ \W_{\rzf}^{(\ell)}(\hat{\H}^{(\ell)}, \alpha^{(n)}) \big\}_{\ell=n+1}^{N} \Big) \Big] \bigg]
\end{align}
\hrulefill
\vspace{-1mm}
\end{figure*}

\subsection{Deterministically Hierarchical D-CSI Model} \label{sec:DCSI_hier}

{\narrowstyleb In this paper, we analyze a deterministically hierarchical network MIMO system, whereby an order is established among the TXs in such a way that TX~$n$ has access not only to $\hat{\H}^{(n)}$ but also to $\{ \hat{\H}^{(\ell)} \}_{\ell=1}^{n-1}$. Hence, such a deterministically hierarchical D-CSI structure allows each TX to determine exactly the strategies computed by the less informed TXs; nevertheless, the strategies used by the more informed TXs can only be predicted imperfectly since their channel estimates are not known.}

In this setting, we have a team decision problem where each TX~$n$ computes its precoding submatrix with the objective of maximizing the ergodic sum rate given the local channel estimate $\hat{\H}^{(n)}$ and the precoding submatrices computed by~the~less informed TXs, as shown in \eqref{eq:W_star} at the top of the previous~page.

\vspace{1mm}

\section{Regularized Zero Forcing Precoding} \label{sec:RZF}

We assume that RZF precoding \cite{Pee05} is adopted at each TX. The RZF precoding submatrix used by TX~$n$ has the form\footnote{Note that this way of enforcing the per-TX power normalization is not necessarily optimal; however, it does not require any additional information exchange between the TXs.} \addtocounter{equation}{-5}
\begin{align}
\nonumber & \hspace{-2mm} \W_{\rzf}^{(n)}(\hat{\H}^{(n)}, \alpha^{(n)}) \triangleq \sqrt{P_{n}} \\
\label{eq:W_rzf} & \hspace{-2mm} \times \frac{\Deltab_{n}^{\tran} \hat{\H}^{(n)} \big( (1 - \alpha^{(n)}) (\hat{\H}^{(n)})^{\herm} \hat{\H}^{(n)} + \alpha^{(n)} \I_{K} \big)^{-1}}{\big\| \Deltab_{n}^{\tran} \hat{\H}^{(n)} \big( (1 - \alpha^{(n)}) (\hat{\H}^{(n)})^{\herm} \hat{\H}^{(n)} + \alpha^{(n)} \I_{K} \big)^{-1} \big\|_{\rmF}}
\end{align}
where $\Deltab_{n} \triangleq [\0_{M_{n} \times \sum_{\ell=1}^{n-1} M_{\ell}} \ \I_{M_{n}} \ \0_{M_{n} \times \sum_{\ell=n+1}^{N} M_{\ell}}]^{\tran} \in \Compl^{M \times M_{n}}$ is a block selection matrix and $\alpha^{(n)} \in [0,1]$ is the regularization factor. The advantage of RZF precoding stems from the fact that only a one-dimensional real parameter, i.e., the regularization factor, needs to be optimized at each TX~$n$ instead of a $(M_{n} \times K)$-dimensional complex matrix.

With RZF precoding, we have a team decision problem where each TX~$n$ computes its regularization factor with the objective of maximizing the ergodic sum rate given the local channel estimate $\hat{\H}^{(n)}$ and the precoding submatrices computed by the less informed TXs, as shown in \eqref{eq:alpha_star} at the top of the page. In the following, we refer to \eqref{eq:alpha_star} as \textit{optimal approach}. By comparing \eqref{eq:W_star} and \eqref{eq:alpha_star}, it is straightforward to note that adopting RZF at the TXs greatly reduces the complexity of the computation of the precoding submatrices.

\subsection{Lower-Complexity Algorithms} \label{sec:RZF_alg}

Deriving $\alpha_{\star,\rmh}^{(n)}$ as in \eqref{eq:alpha_star} is still impractical due to the expectation over the channel estimates at the more informed TXs conditioned on the channel estimate at TX~$n$, i.e., $\{ \hat{\H}^{(\ell)} \}_{\ell=n+1}^{N} | \hat{\H}^{(n)}$, and the maximization at the objective over the regularization factors of the more informed TXs, i.e., $\{ \alpha^{(\ell)} \}_{\ell=n+1}^{N}$, within the aforementioned expectation. Hence, in the following, we present three suboptimal approaches with the aim of reducing the complexity in the computation of the regularization factors. The algorithms are presented in increasing order of both performance and computational~complexity.
\begin{itemize}
\item[-] \textit{Naive approach.} Each TX~$n$ assumes that its local CSI is perfect and shared by the more informed TXs, i.e., $\{ \hat{\H}^{(\ell)} = \H \}_{\ell=n}^{N}$: in this setting, $\alpha_{\star,\rmh}^{(n)}$ simplifies to $\alpha_{\na,\rmh}^{(n)}$ in \eqref{eq:alpha_na}, shown at the top of the page.
\item[-] \textit{Locally robust approach.} Each TX~$n$ assumes that its local CSI is imperfect and shared by the more informed TXs, i.e., $\{ \hat{\H}^{(\ell)} = \hat{\H}^{(n)} \}_{\ell=n+1}^{N}$: in this setting, $\alpha_{\star,\rmh}^{(n)}$ simplifies to $\alpha_{\lr,\rmh}^{(n)}$ in \eqref{eq:alpha_lr}, shown at the top of the page.
\item[-] \textit{Globally robust approach.} Each TX~$n$ assumes that its~local CSI is imperfect and not shared by the more informed TXs; however, in order to reduce the computational complexity with respect to \eqref{eq:alpha_star}, it neglects the possibly different regularization factors adopted by the more informed TXs: in this setting, $\alpha_{\star,\rmh}^{(n)}$ simplifies to $\alpha_{\gr,\rmh}^{(n)}$ in \eqref{eq:alpha_gr}, shown at the top of the page.
\end{itemize}

In the naive approach \eqref{eq:alpha_na}, neither the local nor the global CSI imperfections are taken into account. On the other hand, compared with the latter, the locally robust approach \eqref{eq:alpha_lr} is robust to local CSI imperfections, although it does not deal with the possibly different channel estimates at the more informed TXs. Lastly, the globally robust approach \eqref{eq:alpha_gr} adds some global robustness to the local robustness of \eqref{eq:alpha_lr} but, unlike \eqref{eq:alpha_star}, it does not involve the maximization of the objective over $\{ \alpha^{(\ell)} \}_{\ell=n+1}^{N}$ within the expectation over $\{ \hat{\H}^{(\ell)} \}_{\ell=n+1}^{N} | \hat{\H}^{(n)}$.

\section{The Case of 2 TXs} \label{sec:2TXS}

In this section, we consider a simple network MIMO system with $N=2$ TXs, where TX~2 has perfect CSI, i.e., $\hat{\H}^{(2)} = \H$. In this scenario, the regularization coefficients at TX~1 and TX~2 are computed building on both the optimal and the lower-complexity approaches \eqref{eq:alpha_star}--\eqref{eq:alpha_gr} as follows.

\begin{itemize}
\item[-] Optimal approach. (cf. \eqref{eq:alpha_star}): \vspace{-1mm} \addtocounter{equation}{+4}
\begin{align}
\nonumber \hspace{-7mm} \alpha_{\star,\rmh}^{(1)} & = \argmax_{\alpha^{(1)} \in [0,1]} \Exp_{\H | \hat{\H}^{(1)}} \bigg[ \max_{\alpha^{(2)} \in [0,1]} R \big( \H, \\
& \hspace{19mm} \W_{\rzf}^{(1)}(\hat{\H}^{(1)}, \alpha^{(1)}), \W_{\rzf}^{(2)}(\H, \alpha^{(2)}) \big) \bigg], \\
\hspace{-7mm} \alpha_{\star,\rmh}^{(2)} & = \argmax_{\alpha^{(2)} \in [0,1]} R \big( \H, \W_{\rzf}^{(1)}(\hat{\H}^{(1)}, \alpha_{\star,\rmh}^{(1)}), \W_{\rzf}^{(2)}(\H, \alpha^{(2)}) \big).
\end{align}
\item[-] Naive approach (cf. \eqref{eq:alpha_na}): \vspace{-1mm}
\begin{align}
\nonumber \hspace{-7mm} \alpha_{\na,\rmh}^{(1)} & = \argmax_{\alpha^{(1)} \in [0,1]} R \big( \hat{\H}^{(1)}, \W_{\rzf}^{(1)}(\hat{\H}^{(1)}, \alpha^{(1)}), \\
& \hspace{43mm} \W_{\rzf}^{(2)}(\hat{\H}^{(1)}, \alpha^{(1)}) \big), \\
\hspace{-7mm} \alpha_{\na,\rmh}^{(2)} & = \argmax_{\alpha^{(2)} \in [0,1]} R \big( \H, \W_{\rzf}^{(1)}(\hat{\H}^{(1)}, \alpha_{\na,\rmh}^{(1)}), \W_{\rzf}^{(2)}(\H, \alpha^{(2)}) \big).
\end{align}
\item[-] Locally robust approach (cf. \eqref{eq:alpha_lr}): \vspace{-1mm}
\begin{align}
\nonumber \hspace{-7mm} \alpha_{\lr,\rmh}^{(1)} & = \argmax_{\alpha^{(1)} \in [0,1]} \Exp_{\H | \hat{\H}^{(1)}} \Big[ R \big( \H, \W_{\rzf}^{(1)}(\hat{\H}^{(1)}, \alpha^{(1)}), \\
& \hspace{42mm} \W_{\rzf}^{(2)}(\hat{\H}^{(1)}, \alpha^{(1)}) \big) \Big], \\
\hspace{-7mm} \alpha_{\lr,\rmh}^{(2)} & = \argmax_{\alpha^{(2)} \in [0,1]} R \big( \H, \W_{\rzf}^{(1)}(\hat{\H}^{(1)}, \alpha_{\lr,\rmh}^{(1)}), \W_{\rzf}^{(2)}(\H, \alpha^{(2)}) \big).
\end{align}
\item[-] Globally robust approach (cf. \eqref{eq:alpha_gr}): \vspace{-1mm}
\begin{align}
\nonumber \hspace{-7mm} \alpha_{\gr,\rmh}^{(1)} & = \argmax_{\alpha^{(1)} \in [0,1]} \Exp_{\H | \hat{\H}^{(1)}} \Big[ R \big( \H, \W_{\rzf}^{(1)}(\hat{\H}^{(1)}, \alpha^{(1)}), \\
& \hspace{46mm} \W_{\rzf}^{(2)}(\H, \alpha^{(1)}) \big) \Big], \\
\hspace{-7mm} \alpha_{\gr,\rmh}^{(2)} & = \argmax_{\alpha^{(2)} \in [0,1]} R \big( \H, \W_{\rzf}^{(1)}(\hat{\H}^{(1)}, \alpha_{\gr,\rmh}^{(1)}), \W_{\rzf}^{(2)}(\H, \alpha^{(2)}) \big).
\end{align}
\end{itemize}

\vspace{-2mm}

\subsection{Numerical Results} \label{sec:2TXS_num}

In the above setting, we provide numerical results with the purpose of: \textit{i)} evaluating the gains brought by the deterministically hierarchical D-CSI configuration over the classical non-hierarchical D-CSI setup; and \textit{ii)} assessing the performance of the proposed lower-complexity algorithms. Under the assumption of uniform linear arrays (ULAs) at the TXs, the channel covariance matrices are constructed using the angle spread model (see, e.g., \cite{Yin13}): hence, for each RX $k$, we have $\Sigmab_{k} = \blkdiag (\Sigmab_{k1}, \Sigmab_{k2})$ with $\Sigmab_{kn} = \beta_{kn}^{2} \Exp [\a(\theta_{kn}) \a^{\herm}(\theta_{kn})]$, where $\beta_{kn} > 0$ is the average attenuation of $\h_{kn}$, $\a(\theta) \in \Compl^{Mn \times 1}$ is the steering vector given by
\begin{align}
\a(\theta) \triangleq \big[ 1 \ e^{-j 2 \pi \delta \cos \theta} \ e^{-j 2 \pi (M_{n} - 1) \delta \cos \theta}]
\end{align}
with $\delta$ being the ratio between the antenna spacing at the TX and the signal wavelength, and $\theta_{kn} \in [0, \pi]$ represents the random angle of departure (AoD) between TX~$n$ and RX~$k$. Without loss of generality, we fix $\delta = 0.5$ and assume a uniform distribution for the AoDs such that $\theta_{kn} \in [\bar{\theta}_{kn} - \Delta \theta, \bar{\theta}_{kn} + \Delta \theta]$, where $\bar{\theta}_{kn}$ is the average AoD between TX~$n$ and RX~$k$ and $\Delta \theta$ denotes the angle spread.
We examine a setup with the two TXs facing each other at a distance of $d=40$~m and $K=5$ angularly equispaced RXs in $[\pi/4, 3\pi/4]$ between the TXs. Moreover, we consider $\beta_{kn} = r_{k n}^{-\eta/2}$, where $r_{k n}$ represents the distance between TX~$n$ and RX~$k$ and $\eta = 2$ is the pathloss exponent. Lastly, we assume $M_{1} = M_{2} =4$ for the number of transmit antennas, $\Delta \theta = \pi/8$ for the angle spread, $\sigma^{2} = 0$~dBm for the noise power at the RXs, and $\{ \Upsilonb^{(n)} = \I \}_{n=1}^{N}$ for the error covariance matrices at the TXs.


\begin{figure}[t!]
\centering
\includegraphics[scale=0.7]{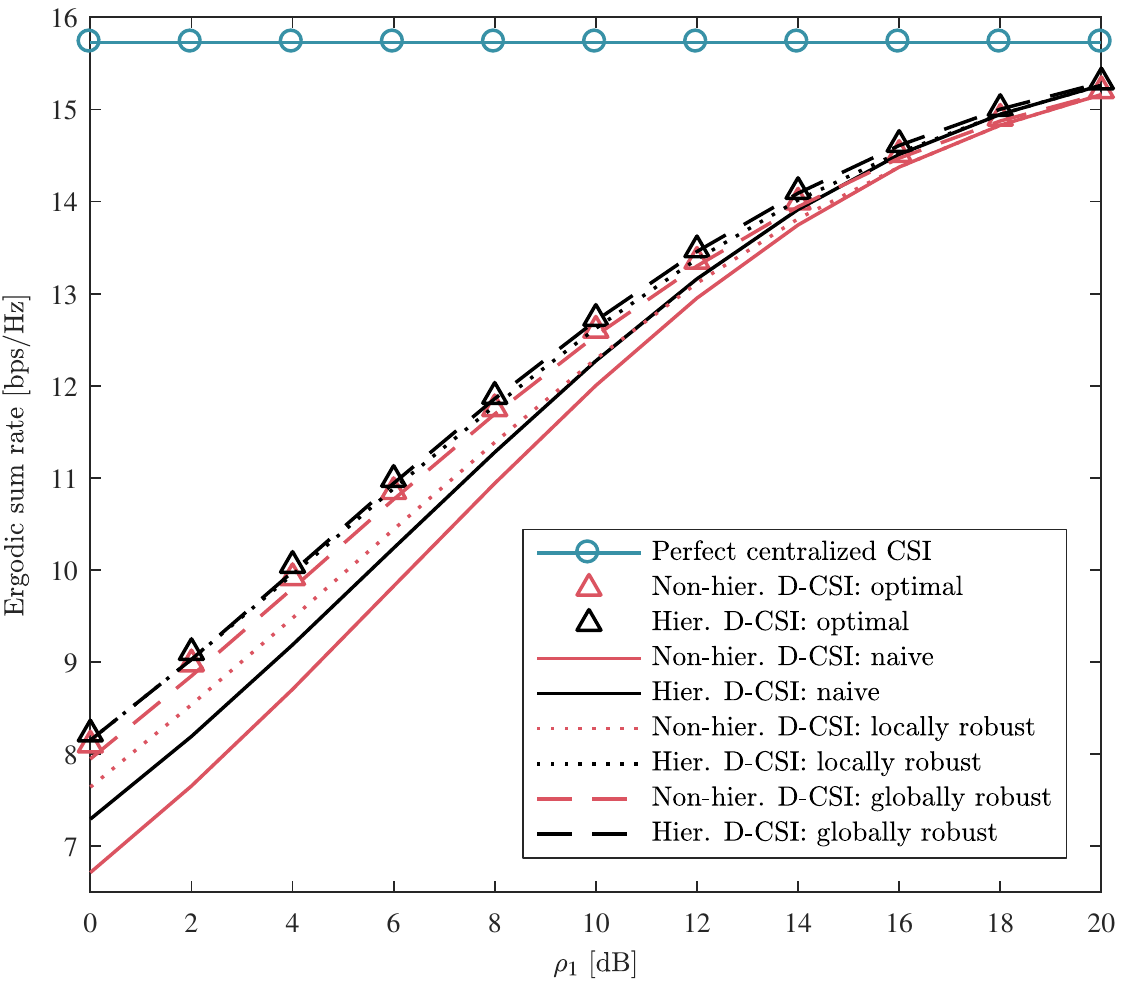} \vspace{-2mm}
\caption{Ergodic sum rate VS feedback SNR of TX~1, with $P_{1} = P_{2} = 10$~dBW.}
\label{fig:rate_VS_rho1} \vspace{-4mm}
\end{figure}

\begin{figure}[t!]
\centering
\includegraphics[scale=0.70]{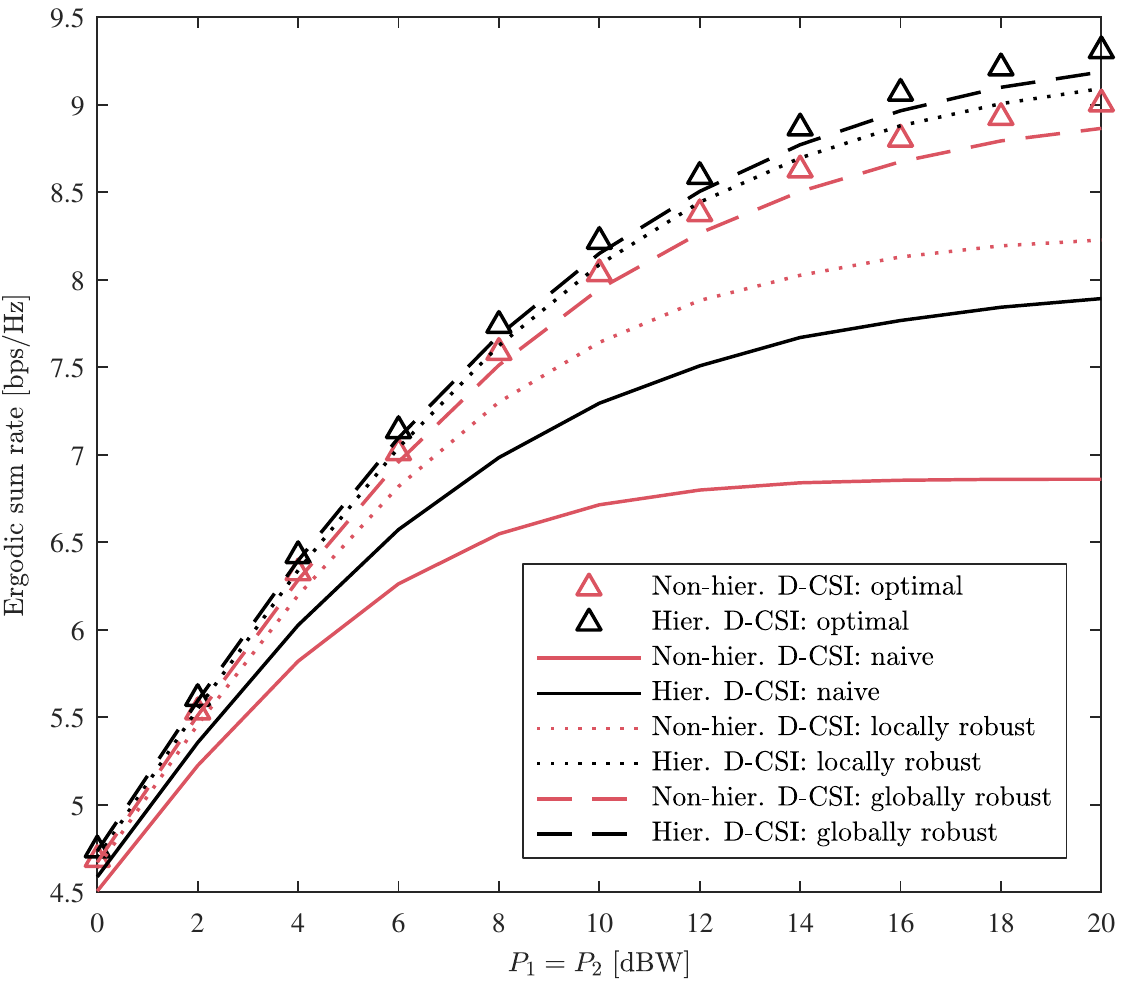} \vspace{-2mm}
\caption{Ergodic sum rate VS per-TX power constraint, with $\rho_{1} = 0$~dB.}
\label{fig:rate_VS_Pn} \vspace{-4mm}
\end{figure}

{\narrowstyleb We analyze the ergodic sum rate as performance metric, which is computed via Monte Carlo simulations with $10^{3}$ realizations of the channel $\H$ and of the channel estimate at TX~1 $\hat{\H}^{(1)}$. Figure~\ref{fig:rate_VS_rho1} considers $P_{1} = P_{2} = 10$~dBW and plots the ergodic sum rate against the quality of the channel estimation expressed in terms of feedback SNR at TX~1, defined as $\rho_{1} \triangleq (1 - \epsilon_{1}^{2})/\epsilon_{1}^{2}$ (cf. \eqref{eq:csi}); here, the performance obtained with perfect and centralized CSI is also included for comparison. First of all, the hierarchical D-CSI setting always outperforms the non-hierarchical D-CSI counterpart in terms of ergodic rate; furthermore, in this hierarchical D-CSI setup, the locally robust approach nearly achieves the performance of the globally robust one, which is in turn remarkably close to the optimal RZF strategy (note that the same does not hold for non-hierarchical D-CSI).
The performance gap between the different algorithms appears more evident from Figure~\ref{fig:rate_VS_Pn}, which illustrates the ergodic sum rate against the per-TX power constraint with $\rho_{1} = 0$~dB. It is straightforward to observe that the overall performance gain brought by the hierarchical D-CSI structure becomes larger as the transmit power increases.}

\subsection{Information Exchange VS Cooperation Gain}  \label{sec:2TXS_pc}

In this section, we briefly address the following question: \textit{what is the cost of the information exchange implied by the deterministically hierarchical D-CSI structure?} Suppose that TX~2 receives only a quantized version of the precoding submatrix $\W_{\rzf}^{(1)}(\hat{\H}^{(1)}, \alpha^{(1)})$ from TX~1 via an out-of-band single-input single-output (SISO) feedback channel with bandwidth $B$, and that the per-TX power budget $P_{1}$ has to accommodate both feedback and downlink transmission: in this regard, we use $P_{1,\fb}$ and $P_{1,\tx}$ to denote the feedback and transmit power, respectively, with $P_{1,\fb} + P_{1,\tx} = P_{1}$. Note~that sharing $\W_{\rzf}^{(1)}(\hat{\H}^{(1)}, \alpha^{(1)})$ rather than $\hat{\H}^{(1)}$ is preferable not only to reduce the information exchange (the former is a $(M_{1} \times K)$-dimensional matrix whereas the latter is $(M \times K)$-dimensional) but also to ease the computational burden at~TX~2.

\begin{figure}[t!]
\centering
\includegraphics[scale=0.70]{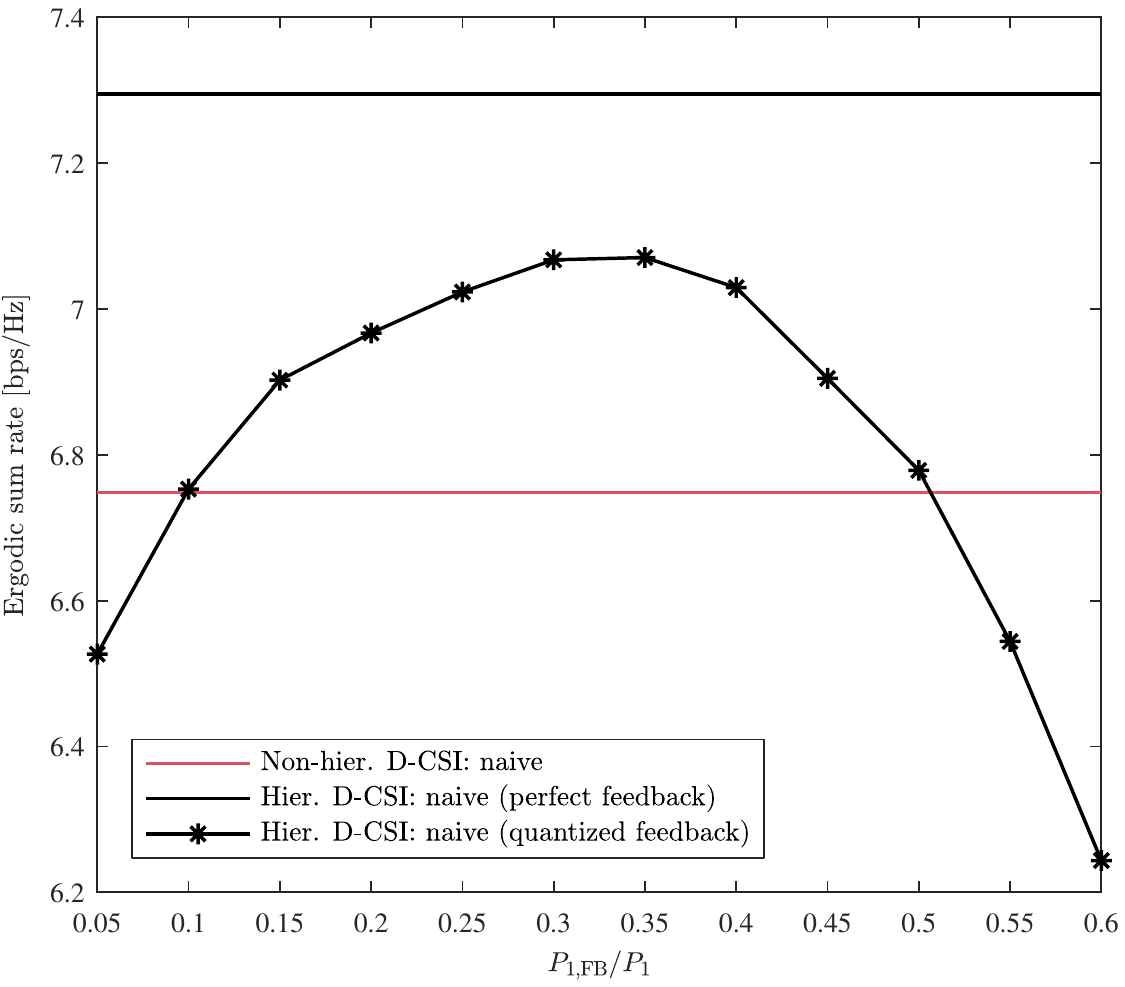} \vspace{-2mm}
\caption{Ergodic sum rate VS feedback power, with $P_{1} = P_{2} = 10$~dBW, $P_{1,\tx} = 10 \log_{10}(P_{1} - P_{1,\fb})$~dBW, and $\rho_{1} = 0$~dB.}
\label{fig:rate_VS_Pfb} \vspace{-4mm}
\end{figure}

Denoting by $\xi$ the number of feedback bits that can be transmitted by TX~1 during the coherence time $T$, let us assume that the two TXs share a common codebook $\setW \triangleq \{ \hat{\W}_{q}^{(1)} \}_{q=1}^{2^{\xi}}$, where each matrix $\hat{\W}_{q}^{(1)} \in \Compl^{(M \times K)}$ has $\| \hat{\W}_{q}^{(1)} \|_{\rmF}^{2} = P_{1}$: then, TX~1 computes $\W_{\rzf}^{(1)}(\hat{\H}^{(1)}, \alpha^{(1)})$ and sends the index
\begin{align}
\hat{q} \triangleq \argmin_{q \in [1, 2^{\xi}]} \| \hat{\W}_{q}^{(1)} - \W_{\rzf}^{(1)} (\hat{\H}^{(1)}, \alpha^{(1)}) \|_{\rmF}
\end{align}
to TX~2, and $\sqrt{P_{1,\tx}} \hat{\W}_{q}^{(1)}$ is taken into account by the latter for the computation of $\W_{\rzf}^{(2)}(\H, \alpha^{(2)})$. Hence, the number of feedback bits is determined by the feedback power as
\begin{align}
\xi \triangleq \left\lfloor B T \log_{2} \bigg( 1 + d^{-\eta} \frac{P_{1,\fb}}{\sigma^{2}} \bigg) \right\rfloor.
\end{align}
{\narrowstyleb Assuming $B=1$~kHz and $T=5$~ms, Figure~\ref{fig:rate_VS_Pfb} compares the hierarchical D-CSI setup} with the non-hierarchical D-CSI counterpart using the naive approach. Interestingly, the former outperforms the latter for $P_{1,\fb}/P_{1} \in [0.05, 0.5]$; more specifically, the ergodic sum rate is maximized when approximately 35\% of the power budget is dedicated to the feedback.

\section{Conclusions} \label{sec:CONCL}

Enforcing a hierarchical information structure is a promising solution to boost the performance of network MIMO systems in presence of distributed channel state information (D-CSI). In this paper, we formulate the general joint precoding optimization problem under D-CSI as a team decision problem and particularize it to a deterministically hierarchical D-CSI scenario. Imposing a specific structure on the precoding matrices, based on regularized zero forcing (RZF) precoding, we propose naive, locally robust, and globally robust suboptimal strategies for the joint precoding design. Focusing on the simple case of two TXs, we show that the deterministically hierarchical D-CSI setup yields significant gains over the classical non-hierarchical D-CSI counterpart (larger gains are expected for {\narrowstyleb a higher number of TXs) and that the locally and globally robust approaches perform remarkably close to the optimal RZF strategy.}

\addcontentsline{toc}{chapter}{References}
\bibliographystyle{IEEEtran}
\bibliography{IEEEabrv,refs_itatz}

\begin{thebibliography}{10}
\providecommand{\url}[1]{#1}
\csname url@samestyle\endcsname
\providecommand{\newblock}{\relax}
\providecommand{\bibinfo}[2]{#2}
\providecommand{\BIBentrySTDinterwordspacing}{\spaceskip=0pt\relax}
\providecommand{\BIBentryALTinterwordstretchfactor}{4}
\providecommand{\BIBentryALTinterwordspacing}{\spaceskip=\fontdimen2\font plus
\BIBentryALTinterwordstretchfactor\fontdimen3\font minus
  \fontdimen4\font\relax}
\providecommand{\BIBforeignlanguage}[2]{{%
\expandafter\ifx\csname l@#1\endcsname\relax
\typeout{** WARNING: IEEEtran.bst: No hyphenation pattern has been}%
\typeout{** loaded for the language `#1'. Using the pattern for}%
\typeout{** the default language instead.}%
\else
\language=\csname l@#1\endcsname
\fi
#2}}
\providecommand{\BIBdecl}{\relax}
\BIBdecl

\bibitem{Ges10}
D.~Gesbert, S.~Hanly, H.~Huang, S.~Shamai~(Shitz), O.~Simeone, and W.~Yu,
  ``Multi-cell {MIMO} cooperative networks: {A} new look at interference,''
  \emph{{IEEE} J. Sel. Areas Commun.}, vol.~28, no.~9, pp. 1380--1408, Dec.
  2010.

\bibitem{deK12}
P.~de~Kerret and D.~Gesbert, ``Degrees of freedom of the network {MIMO} channel
  with distributed {CSI},'' \emph{{IEEE} Trans. Inf. Theory}, vol.~58, no.~11,
  pp. 6806--6824, Nov. 2012.

\bibitem{Moz15}
M.~Mozaffari, W.~Saad, M.~Bennis, and M.~Debbah, ``Drone small cells in the
  clouds: Design, deployment and performance analysis,'' in \emph{Proc. {IEEE}
  Global Commun. Conf. (GLOBECOM)}, San Diego, CA, USA, Dec. 2015.

\bibitem{Rad62}
R.~Radner, ``Team decision problems,'' \emph{Ann. Math. Statist.}, vol.~33,
  no.~3, 1962.

\bibitem{She08}
M.~B. Shenouda and T.~N. Davidson, ``On the design of linear transceivers for
  multiuser systems with channel uncertainty,'' \emph{{IEEE} J. Sel. Areas
  Commun.}, vol.~26, no.~6, pp. 1015--1024, Aug. 2008.

\bibitem{Vuc09}
N.~Vucic, H.~Boche, and S.~Shi, ``Robust transceiver optimization in downlink
  multiuser {MIMO} systems,'' \emph{{IEEE} Trans. Signal Process.}, vol.~57,
  no.~9, pp. 3576--3587, Sept. 2009.

\bibitem{deK16}
P.~de~Kerret and D.~Gesbert, ``Quantized team precoding: A robust approach for
  network {MIMO} under general {CSI} uncertainties,'' in \emph{Proc. {IEEE}
  Int. Workshop Signal Process. Adv. in Wireless Commun. (SPAWC)}, Edinburgh,
  UK, July 2016.

\bibitem{Lak16}
T.~R. Lakshmana, A.~T\"{o}lli, and T.~Svensson, ``Improved local precoder
  design for {JT-CoMP} with periodical backhaul {CSI} exchange,'' \emph{{IEEE}
  Commun. Lett.}, vol.~20, no.~3, pp. 566--569, Mar. 2016.

\bibitem{deK14}
P.~de~Kerret, R.~Fritzsche, D.~Gesbert, and U.~Salim, ``Robust precoding for
  network {MIMO} with hierarchical {CSIT},'' in \emph{Proc. {IEEE} Int. Symp.
  Wireless Commun. Syst. (ISWCS)}, Barcelona, Spain, Aug. 2014.

\bibitem{Pee05}
C.~B. Peel, B.~M. Hochwald, and A.~L. Swindlehurst, ``A vector-perturbation
  technique for near-capacity multiantenna multiuser
  communication---{P}art~{I}: {C}hannel inversion and regularization,''
  \emph{{IEEE} Trans. Commun.}, vol.~53, no.~1, pp. 195--202, Jan. 2005.

\bibitem{Yin13}
H.~Yin, D.~Gesbert, M.~Filippou, and Y.~Liu, ``A coordinated approach to
  channel estimation in large-scale multiple-antenna systems,'' \emph{{IEEE} J.
  Sel. Areas Commun.}, vol.~31, no.~2, pp. 264--273, Feb. 2013.

\end{thebibliography}

\end{document}